\documentclass{article}

\usepackage{PRIMEarxiv}

\usepackage[utf8]{inputenc}
\usepackage[T1]{fontenc}
\usepackage{hyperref}
\usepackage{url}
\usepackage{booktabs}
\usepackage{amsfonts}
\usepackage{nicefrac}
\usepackage{microtype}
\usepackage{lipsum}
\usepackage{graphicx}

\usepackage{amsmath, bm}
\usepackage{amssymb}
\usepackage{xcolor}
\usepackage{longtable}
\usepackage{natbib}

\newcommand{\Sstar}{S^{\star}}
\newcommand{\E}{\mathbb{E}}
\newcommand{\Var}{\mathrm{Var}}
\newcommand{\dskl}{d_{\mathrm{SKL}}}

\title{Statistical Detectability of Coordination}

\author{
  Letizia Iannucci \\
  Aalto University \\
  Finland \\
   \And
  Antonis Matakos \\
  Stockholm University \\
  Sweden\\
     \And
  Mikko Kivelä \\
  Aalto University \\
  Finland\\
}

\makeatletter
\newcommand{\titlefootnotesetup}{%
  \renewcommand{\thefootnote}{\fnsymbol{footnote}}%
  \renewcommand{\@makefnmark}{\hbox to \z@{$^{\@thefnmark}$\hss}}%
  \long\def\@makefntext##1{%
    \parindent 1em\noindent
    \hbox to 1.8em{\hss $\m@th ^{\@thefnmark}$}##1%
  }%
}
\newcommand{\twocolumnmaketitle}[2]{%
  \twocolumn[{%
    \begingroup
      \titlefootnotesetup
      \@maketitle
    \endgroup
    \begin{abstract}
    Information operations exploit coordinated activity on social media to amplify messages and manipulate public discourse. A common approach to coordination detection is to project account-entity interactions into account similarity networks based on temporal proximity, action counts, or behavioural similarity, and identify dense subgraphs or behavioural clusters. However, the current methodology finds groups of co-activated accounts or partitions all users into most similarly acting clusters, even when the accounts are acting independently of each other or the co-activity is due to some expected source. We formulate coordination detection based on hypothesis testing of account pairs and compare different null models. We connect the resulting pairwise scores to the planted-submatrix detection problem and derive statistical and computational limits as functions of group size, population size, and signal strength. We estimate the signal for 7 modalities of coordination over 26 labelled information-operation datasets, and compare the predicted regime with the performance of recovery heuristics. Our method achieves higher agreement with the labelled ground truth than previously reported, but the choice of null model is important and it changes the regime of most modalities. While it is always possible to extract a set of significantly densely co-acting accounts across nearly all modalities regardless of regime, density alone is uninformative of detection performance, rendering anomalous cluster discovery a poor proxy for recovering an information operation. Detectability boundaries establish whether an operation of a given size and signal can be recovered at all, and how much confidence a flagged group warrants.
    \end{abstract}
    \keywords{#2}
    \vspace{1.5em}
  }]%
  \begingroup
    \titlefootnotesetup
    \@thanks
  \endgroup
  \thispagestyle{empty}%
  \setcounter{footnote}{0}%
  \let\@thanks\@empty
}
\makeatother

\begin{document}

\twocolumnmaketitle
{}
{coordinated inauthentic behaviour \and information operations \and hypothesis testing \and planted submatrix}

\section{Introduction}

Information operations exploit coordinated activity on social media to manipulate conversations and spread disinformation through synchronised actions, such as simultaneous hashtag usage, coordinated replying, or orchestrated content amplification \citep{starbird2019disinformation, ng2022combined, pacheco2021uncovering, ferrara2016rise}. Detecting such orchestrated campaigns from their online traces is critical to preserve the integrity of online spaces.

To isolate coordinated groups, detection pipelines typically build a coordination network, linking accounts whose actions resemble one another or co-occur close in time, such as accounts posting the same URL, reusing the same hashtag, reposting the same post, or replying in near-synchrony \citep{pacheco2021uncovering, nizzoli2021coordinated, weber2021amplifying, sharma2021identifying, cima2024coordinated, mannocci2025multimodal}. Clustering of account behaviour or community detection methods then recover groups whose internal connections are denser than their connections to the rest of the network or whose activity patterns are similar \citep{magelinski2022synchronized, luceri2024unmasking, mannocci2026detection}. Parameters are typically tuned empirically or set from prior domain knowledge, treating ``similar enough,'' ``close enough,'' or ``dense enough'' as a design choice. However, the choice of time window and what kind of clusters count as synchronised determine which coordination is detectable \citep{panayiotou2026setting, mannocci2025multimodal}.

A central challenge is to separate coordination from co-activity that arises by chance or from patterns expected among organic users. For example, users can act close in time by reacting to the same trending topic or hashtag, by following their own daily rhythms, or simply because they are very active. Accounting for such activity can change conclusions: \citet{pante2025beyond} show that, once organic control accounts are included, there is no statistical evidence of the inter-state coordination that earlier studies had suggested. We argue that a detected group is evidence of coordination only if expected patterns and resulting chance co-activity cannot explain its total signal. Benchmarks on labelled data show whether a detector recovers known operations, but do not quantify how surprising the observed structure is under the assumption of independent behaviour.

We treat coordination as a hypothesis-testing problem with two levels. At the level of pairs, we test each pair of accounts against a null model with prescribed patterns but otherwise independent activity, and turn its $p$-value into a score, so that a coordinating group of $k$ accounts among $n$ becomes a $k \times k$ block of elevated scores in an $n \times n$ matrix. At the level of groups, we reduce coordination detection to the planted submatrix problem, for which the conditions under which such a block can be detected are known \citep{butucea2013detection, ma2015computational, cai2017computational, brennan2019universality, brennan2020reducibility}. In contrast to the typical approach of graph clustering without a statistical model, which always returns a partition of the graph and where quality functions such as modularity are well-known to be prone to overfitting \cite{guimera2004modularity, peixoto2023descriptive}, this formulation allows us to answer under what conditions coordinated behaviour can be detected. We derive a phase diagram with three detection regimes (impossible, hard, and easy), whose boundaries are functions of group size, population size, and signal strength. We also provide a theoretical basis for the recent empirical observations that multi-modality improves coordination detection \citep{mannocci2025multimodal, iannucci2026detecting}, showing that combining $M$ independent modalities of equal signal lowers every detection threshold by a factor of $1/\sqrt{M}$, so an operation that cannot be detected in any single layer can become detectable. 

We measure the signal strength of 26 labelled information operations \citep{seckin2025labeled} and compare the predicted regime with the recovery of two greedy heuristics and a spectral method on 7 modalities and an aggregated view of each operation. Since chance synchrony has several sources, we implement four null models, and measure how much the conclusions depend on the definition of chance. The measured signal depends on the null model, because each null model attributes different parts of the observed synchrony to chance. 
Our approach has several assumptions that are common in the literature, but are often implicit, and made explicit by our principled statistical framework. This allows us to test the effect of these various assumptions to the peformance of the coordination detection.

This paper makes several contributions by formulating coordination detection as hypothesis testing within the planted-submatrix problem. First, we derive information-theoretic and computational boundaries for the detection of a coordinated campaign. Second, we introduce a new methodology for coordination detection, which achieves higher agreement with the labelled ground truth than previously reported \citep{iannucci2026detecting}, although the evaluation protocols cannot be made exactly equivalent. Third, we evaluate four null models on 26 information operations and show how and why the predicted regimes depend on the choice of the null model. Fourth, we show that multi-modal analysis lowers detection thresholds by $1/\sqrt{M}$. Finally, we show that a permutation test on the density of a returned set does not indicate recovery: the heuristics return significantly dense sets even when they do not recover the operation. Thus, structural properties of the coordination networks alone are an insufficient proxy for coordination detection.

\section{Related Work}
\label{sec:related}

\textbf{Coordination detection.} Most methods for coordination detection build a network in which an edge links two accounts that act alike, and extract groups from the resulting network by connected components, community detection, or clustering \citep{pacheco2021uncovering, nizzoli2021coordinated, magelinski2022synchronized, cima2024coordinated}. Such pipelines have been applied to elections \citep{nizzoli2021coordinated, tardelli2024multifaceted}, to information operations \citep{cima2024coordinated, luceri2024unmasking}, and across platforms \citep{cinus2025exposing}. The modality, the size of the time window, and the threshold to retain significant edges are usually design choices \citep{mannocci2026detection}. The clustering and community detection methods used in this step return a partition or a group for any input network, including random graphs. Commonly used quality metrics such as modularity are known to overfit: modularity can take high values even on random graphs \citep{guimera2004modularity}, so a high modularity is not evidence of communities \citep{peixoto2023descriptive}. The question is therefore not only whether a coordinating cluster exists, but whether its members co-act more than independent accounts would. A few studies test suspicious accounts against control accounts \citep{pante2025beyond, mungari2026temporal}, or compare the coordination of accounts in an information operation with samples of regular users \citep{keller2020political}. Still, the field lacks a formal and statistically grounded definition of coordination, including well-defined null models \citep{mannocci2026detection}, and there is limited knowledge of when coordination is even detectable.

\textbf{Detecting a planted submatrix.} Like other methods that extract groups of nodes, dense-subgraph extraction returns a dense set for every input, including pure noise \citep{charikar2000greedy, hooi2016fraudar, shabalin2009finding}. Statistical work asks when a planted block can be detected. Scan statistics detect it above the information-theoretic boundary \citep{kulldorff1997spatial, arias2011detection, addario2010combinatorial}, while polynomial-time methods need a larger block or a stronger signal \citep{cai2017computational}. Locating
the block, i.e. recovery, is at least as hard as detecting it \citep{cai2017computational}. Polynomial-time recovery is known above a signal of the order of the spectral boundary \citep{hajek2018submatrix}, and the statistical and computational limits of recovery are also known for several planted submatrices \citep{chen2016statistical}. The computational lower bound holds for broad classes of distributions of the entries of the matrix \citep{brennan2019universality}. To our knowledge, these results have not yet been applied to coordination detection.

\section{Methods}
\label{sec:methods}

\subsection{Co-action layers and null models}
\label{sec:coaction}

\begin{figure}[ht]
    \centering
    \includegraphics[width=\linewidth]{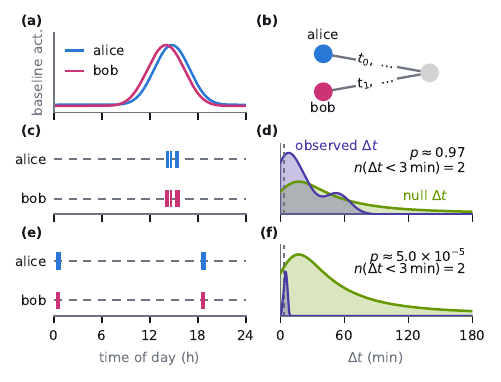}
    \caption{The $p$-value of a pair under the account-activity null model. (a) Daily activity profiles of two accounts. (b) Two users co-act. (c, e) Two patterns of co-action times. (d, f) Distribution of the time intervals $\Delta t$ between co-actions, observed and under the null model, with the $p$-value; the dotted line at $3$ minutes shows that counts over a fixed time window might not distinguish the two co-action patterns.
    }
    \label{fig:nullmodel}
\end{figure}

We model the activity of $n$ accounts. An \emph{item} is the object of an action, and a \emph{modality} is a type of action on an item. The available modalities depend on the social media platform. In Section~\ref{sec:results} we use seven modalities: using a hashtag, sharing a URL, mentioning an account, replying to a post, replying to an account, reposting a post, and reposting an account. Two accounts \emph{co-act} in a modality if both act on the same item. Each modality defines a \emph{layer}, a graph whose edges connect accounts that co-act at least once. Such pairs of accounts are \emph{tested} for temporal synchronisation, and each edge carries a $p$-value $p^{(l)}_{ij}$ that measures how unusually the pair acts on their shared items, under a null model. 

Accounts that coordinate have been observed repeatedly performing similar or identical actions within a short span of each other, more often than other accounts \citep{pacheco2021uncovering, nizzoli2021coordinated, cima2024coordinated}. Making this precise requires a choice of how short ``short'' is, yet different modalities operate on different natural time scales \citep{iannucci2026detecting}, and there is no consensus on a single window value that works across modalities or across documented coordinated campaigns \citep{panayiotou2026setting, mannocci2026detection}.

For a tested pair $(i,j)$ and a shared item $o$, let $T_i(o)$ and $T_j(o)$ denote their respective action timestamps on $o$. For each $t\in T_i(o)$, the interval to the next action of $j$ is $\Delta_{i\to j}(t;o)=\min_{s\in T_j(o),\,s\ge t}(s-t)$, provided such an action exists. We define $\bar\Delta_{ij}(o)$ as the mean of the existing intervals in both directions. This statistic requires no fixed time window.

A null model generates randomised replicates of the actions on $o$. With $B=100$ item-level Monte Carlo replicates, $p_{ij}(o)=(r+1)/(B+1)$, where $r$ counts replicates whose mean interval does not exceed the observed $\bar\Delta_{ij}(o)$. We use Fisher's method to combine the $p$-values of all items shared by the pair in modality $l$ into $p_{ij}^{(l)}$.

We consider four null models that preserve each account's actions on each item but randomize their timestamps to test for synchrony. We can approximately order the null models by the amount of observed temporal structure that they preserve \citep{gauvin2022randomized}. The \emph{uniform-modality null model} draws the time of each action uniformly from an evenly spaced grid with one point per action, between the first and the last action in the modality. The \emph{uniform-account null model} instead uses a separate grid per account, spanning its first and last actions across all modalities, thus preserving each account's active period. Both null models attribute every other timing pattern, including daily rhythms and reactions to the same item, to synchrony. The \emph{account-activity null model} preserves the date of each action and redraws its hour from the hourly profile of the account for that day of the week. Two accounts being active at the same hours is then not by itself evidence of coordination, if it is compatible with their baseline activity profiles. This null model does not model reactions to an item, so it overstates the synchrony of accounts that respond to the same post or URL soon after it appears (Section~\ref{sec:results}). The \emph{item-virality null model} redraws the action times of an account on item $x$ uniformly, with replacement, from the action times of all accounts on $x$. This keeps the adoption curve of each item in expectation, so accounts that react to a viral item at the same time are not by themselves evidence of coordination.

\subsection{Coordination as planted submatrix}
\label{sec:coord-as-planted}

We model coordination as a single group of accounts whose activity produces pairwise independent $p$-values from a coordination distribution $P$, while all  other pairs produce $p$-values from the uniform distribution $Q$ on $(0,1)$. This model defines a hypothesis, $H_1$, which is tested against the null $H_0$ (absence of coordination). Our modeling makes the following two assumptions. First, the $p$-values are independent, treating all the dependencies in the activity signals of nodes as encoded in the null model. Second, we assume that there is no internal structure among the coordinated or uncoordinated nodes. The latter is analogous to the typical, but often implicit, assumption in community detection methods that the clusters are internally homogeneous \cite{fortunato2010community}.

For a coordinating pair of accounts $(i,j)$ we assume $p^{(l)}_{ij}\sim P=\mathrm{Beta}(\lambda,1)$ with $0<\lambda<1$. This is a common family for modelling $p$-values under an alternative hypothesis because its non-increasing density assigns greater probability to small $p$-values \citep{sellke2001calibration}. Moreover, Fisher’s method is optimal for combining independent $p$-values following this family \citep{heard2018choosing}. However, our analysis can be carried out for any other continuous distribution $P$ supported on $(0,1)$. The log-likelihood ratio of $\mathrm{Beta}(\lambda,1)$ against the uniform distribution, evaluated at the observed $p^{(l)}_{ij}$, is $\log\lambda+(1-\lambda)(-\log p^{(l)}_{ij})$, which is increasing in $-\log p^{(l)}_{ij}$. For a fixed set of independent pairs sharing the same $\lambda$, rejecting when the sum of $-\log p^{(l)}_{ij}$ exceeds a critical value is uniformly most powerful: among tests with the same significance level, it has the highest probability of detecting coordination for every $0<\lambda<1$ \citep{lehmann2005testing}. 

This motivates the centred score
\begin{equation}
    Y^{(l)}_{ij} = -\log p^{(l)}_{ij}-1,
    \label{eq:score}
\end{equation}
which has mean $0$ and variance $1$ under $H_0$. Under the Beta alternative, $Y^{(l)}_{ij} \sim\mathrm{Exp}(\lambda)-1$, with mean $\mu:=1/\lambda-1>0$, which we call the \emph{signal strength}.
The $Y^{(l)}_{ij}$ scores depend on the choice of null model (Section~\ref{sec:results}) and define, for each layer $l$, a symmetric $n\times n$ matrix where the diagonal is left unspecified. 

Let $\mathcal S_k$ be the family of all subsets of $k$ accounts. Under $H_1$, the coordinating group $\Sstar$ is drawn uniformly from $\mathcal S_k$. Assuming independence across pairs under $H_0$ and conditional independence given $\Sstar$ under $H_1$, the off-diagonal entries follow for $i<j$: 

\begin{align*}
H_0:\quad &Y^{(l)}_{ij}\sim\mathrm{Exp}(1)-1,\\
H_1:\quad &Y^{(l)}_{ij}\sim
\begin{cases}
\mathrm{Exp}(\lambda)-1, & i,j\in\Sstar,\\
\mathrm{Exp}(1)-1, & \text{otherwise}.
\end{cases}
\end{align*}

Our formulation is a variant of the well-studied planted-submatrix problem, for which regimes of statistical impossibility, computational hardness, and polynomial-time detection have been characterised asymptotically. To derive the corresponding detection boundaries for our setting, we must account for two differences from existing formulations. First, our matrix is symmetric, with $\binom n2$ independent entries and $\binom k2$ independent entries in the planted block. The formulations of \citet{butucea2013detection} and \citet{ma2015computational} allow different sets of planted rows and columns, while \citet{brennan2019universality} consider a single set, as we do, but treat all $k^2$ entries as independent. In contrast, our score matrix is symmetric, with an unspecified diagonal, and therefore has only $\binom n2$ independent entries, of which $\binom k2$ belong to the planted block. 

The computational lower bound of \citet{brennan2019universality} transfers to our problem by a reduction. Given an instance $\mathbf W$ of their problem, we construct $\mathbf Y$ by setting  $Y^{(l)}_{ij}=Y^{(l)}_{ji}=W_{ij}$ for $i<j$. This polynomial-time reduction preserves the distribution under both hypotheses. Thus, a polynomial-time test for our problem would also solve theirs, contradicting the lower bound under the planted clique hypothesis. The reduction transfers the lower bound to our problem, but it does not establish its tightness: the reduction discards $k(k+1)/2$ entries of their block on and below the diagonal. We obtain computational achievability thresholds from the sum and spectral tests, and derive the information-theoretic boundary for our symmetric model in Section~\ref{sec:threeregimes}. Second, unlike the classical Gaussian formulations, our entries follow shifted exponential distributions. We derive the upper bounds from the moment generating function of $Y$ 
and the information-theoretic lower bound from the second moment of the likelihood ratio (Appendix~\ref{app:derivation}).

\subsection{Detection boundaries}
\label{sec:threeregimes}

Next we derive sufficient conditions for detecting the presence of a planted block, using exhaustive search and polynomial-time methods, as functions of the signal strength $\mu$, the total number of accounts $n$, and the block size $k$.

\textbf{Exhaustive search.} The test searches over all possible blocks of size $k$ and rejects $H_0$ when
\[
T^{(l)}_{\text{scan}} = \max_{S\in\mathcal S_k}\sum_{i<j\in S}Y^{(l)}_{ij}
\]
exceeds a critical value $c$, chosen so that the probability of a false detection $\mathbb{P}_0(T_{\mathrm{scan}}>c)\to0$ as $n\to\infty$. We obtain the critical value from a Chernoff bound and a union bound over $\mathcal S_k$ as $c=(1+\varepsilon)\,k\sqrt{(k-1)\log(n/k)}$ for any fixed $\varepsilon>0$ (Appendix~\ref{app:scanthresh}). Under $H_1$, the true coordinating group has expected total score $\binom k2\mu$, so the scan detects the block for $n \to \infty$ when $\mu>\mu_{\mathrm{scan}}=2\sqrt{\log(n/k)/(k-1)}$, as for Gaussian entries \citep{butucea2013detection}.

\textbf{Information-theoretic boundary.} The total error probability of a test is the sum of the probabilities of a false positive under $H_0$ and of a missed detection under $H_1$. Below the information-theoretic boundary, no test can tell $H_1$ from $H_0$ with total error probability below a fixed positive level, regardless of the computational cost \citep{butucea2013detection, ma2015computational, brennan2019universality}. Following the second-moment method \citep{ingster2012nonparametric, addario2010combinatorial, arias2011detection, butucea2013detection, arias2014community}, we bound the total error probability from below through the second moment $\E_0[L^2]$ of the likelihood ratio $L$ of $H_1$ against $H_0$: if $\E_0[L^2]\le\gamma$ for a constant $\gamma>1$, every test has a total error probability of at least $1-\tfrac12\sqrt{\gamma-1}$ \citep{tsybakov2009introduction}. The boundary therefore depends on the error level that one requires, through $\gamma$. We take $\gamma=e$ because with this value the information-theoretic boundary for large groups equals the sum boundary $\mu_{\mathrm{sum}}$ derived below. With this choice, every test has a total error probability of at least $0.34$ below the signal $\mu_{\mathrm{IT}}$ at which $\E_0[L^2]=e$, which is, to leading order (Appendix~\ref{app:it-bound}),
\begin{equation}
    \mu_{\mathrm{IT}}=\min\Big\{\sqrt{2\log(n/k)/(k-1)},\ \sqrt2\,n/k^2\Big\}.
\label{eq:muit-main}
\end{equation}
Another $\gamma$ changes the large-group term $\sqrt2\,n/k^2$ of Equation~\eqref{eq:muit-main} to $\sqrt{2\log\gamma}\,n/k^2$, while the small-group term $\sqrt{2\log(n/k)/(k-1)}$ does not depend on $\gamma$ (Appendix~\ref{app:it-bound}). For small groups, $\mu_{\mathrm{IT}}$ is a factor $\sqrt2$ below the scan boundary $\mu_{\mathrm{scan}}$; for $\mu$ between the two, it is not known whether any test detects the block (Appendix~\ref{app:it-bound}).

\textbf{Computational boundary.} The \emph{computational boundary} is the signal strength above which a polynomial-time test detects the block and below which no polynomial-time test does \citep{ma2015computational, hajek2015computational, brennan2019universality}. As in \citep{ma2015computational} and \citep{brennan2019universality}, we bound it from above with polynomial-time tests. The \emph{sum test} $T^{(l)}_{\mathrm{sum}}=\sum_{i<j}Y^{(l)}_{ij}$ rejects $H_0$ when the sum of all scores is large \citep{butucea2013detection, arias2014community}. Under $H_0$, $T^{(l)}_{\mathrm{sum}}$ has mean $0$ and standard deviation $\sqrt{\binom n2}$; under $H_1$, its mean is $\binom k2\mu$, and near the boundary its standard deviation remains close to $\sqrt{\binom n2}$ (Appendix~\ref{app:sum}). As in \citep{butucea2013detection}, we divide $T^{(l)}_{\mathrm{sum}}$ by its standard deviation under $H_0$. The normalised sum has expected value $0$ under $H_0$ and $\delta=\binom k2\mu\big/\sqrt{\binom n2}$ under $H_1$. For Gaussian entries, the total error of the sum test tends to $0$ when $\delta\to\infty$, and the boundary lies where $\delta$ is approximately one \citep{butucea2013detection}. Following \citep{butucea2013detection}, the sum boundary is the signal strength at which $\delta=1$ (Appendix~\ref{app:sum}):

\[
\mu_{\mathrm{sum}}=\frac{\sqrt{2n(n-1)}}{k(k-1)}.
\]

The \emph{spectral test} uses the largest eigenvalue of the score matrix. Under $H_1$, the block adds to $\mathbf Y$ a matrix with entry $\mu$ on the pairs of the block, whose largest eigenvalue is $\mu(k-1)$; if this eigenvalue is large enough, it raises the largest eigenvalue of $\mathbf Y$ above its value under $H_0$. As in \citep{montanari2015limitation}, the test rejects $H_0$ when the largest eigenvalue of $\mathbf Y/\sqrt n$ exceeds $2+\epsilon$ with $\epsilon > 0$, and it detects the block if and only if $\mu>\mu_{\mathrm{spec}}=\sqrt n/(k-1)$ \citep{montanari2015limitation, knowles2013isotropic} (Appendix~\ref{app:spectral}). 

Since each test detects the block above its own boundary, a polynomial-time test boundary is
\[
\mu>\mu_{\mathrm{comp}}=\min(\mu_{\mathrm{spec}},\mu_{\mathrm{sum}}).
\]
The sum test gives the minimum when $k>\sqrt{2(n-1)}$, and the spectral test otherwise. For $k\ge\sqrt n$, no polynomial-time test detects the block under the planted clique hypothesis when $k^4\,\mathrm{KL}(P\Vert Q)/n^2\to0$ \citep{brennan2019universality}, 
where $\mathrm{KL}(P\Vert Q)$ is the Kullback--Leibler divergence of the distribution $P$ of a block entry from the distribution $Q$ of a null entry. For our scores, $\mathrm{KL}(P\Vert Q)=\mu^2/2\,\big(1+O(\mu)\big)$, so the condition is $\mu=o(\mu_{\mathrm{sum}})$ (Appendix~\ref{app:threeregimes}). For $k<\sqrt n$, the lower bound of \citep{brennan2019universality} applies to our scores only for $\mu$ of order at most $1/\log n$, far below $\mu_{\mathrm{comp}}$ (Appendix~\ref{app:threeregimes}). For $\mu$ between these two values, the bounds do not determine whether a polynomial-time test detects the block.

\textbf{Regimes.} Detection is \emph{impossible} if $\mu<\mu_{\mathrm{IT}}$, \emph{hard} if $\mu_{\mathrm{IT}}\le\mu<\mu_{\mathrm{comp}}$, and \emph{easy} if $\mu\ge\mu_{\mathrm{comp}}$. For $k<\sqrt n$, \emph{hard} thus means that neither the sum test nor the spectral test detects the block. With $k=n^\beta$ and $\mu^2=n^{-\alpha}$, the three boundaries are, to leading order, the lines $\alpha=\beta$ (scan and small-group information-theoretic boundary), $\alpha=2\beta-1$ (spectral), and $\alpha=4\beta-2$ (sum and large-group information-theoretic boundary). Up to terms that vanish for $n \to \infty$, these are the coordinates of \citet{brennan2019universality}. The lines $\alpha=\beta$ and $\alpha=4\beta-2$ meet at $\beta=2/3$: for larger groups the information-theoretic boundary equals the sum boundary, so detection is easy. 

\textbf{Untested pairs.} When two accounts do not co-act, their pair has no $p$-value. We give such pairs the score $0$ under both hypotheses, so the signal strength can still be computed as the mean score over all pairs of the group, tested or not. The boundaries above assume that every pair is tested. With untested entries, the sum of all scores keeps its mean $\binom k2\mu$ under $H_1$ and has a smaller variance under both hypotheses, so the sum test detects the block at a smaller $\mu$ than $\mu_{\mathrm{sum}}$; the same holds for the spectral test when the untested pairs are spread uniformly. The sum and spectral boundaries of a matrix with untested entries are therefore at most the values for a complete matrix. A layer in which $\hat\mu\ge\mu_{\mathrm{comp}}$ is easy, but a layer in which $\hat\mu$ is below $\mu_{\mathrm{IT}}$ or $\mu_{\mathrm{comp}}$ may still lie above the corresponding boundary of its matrix with untested entries, so its regime, impossible or hard, is an approximation.

\textbf{Other distributions for the alternative hypothesis $H_1$.} The assumption $P = \mathrm{Beta}(\lambda,1)$ can be relaxed to any continuous distribution $P$ on $(0,1)$. The scan, sum, and spectral tests detect the block when $\mu=\E_P[Y^{(l)}_{ij}]$ exceeds their boundaries, for any continuous $P$ under which the variance of $Y^{(l)}_{ij}$ stays bounded. The scan rejects $H_0$ when the largest sum of scores over a set of $k$ accounts exceeds its critical value, and the sum test when the sum of all scores exceeds its critical value. Both critical values are computed from the null distribution $\mathrm{Exp}(1)-1$ of the scores, so they do not depend on $P$. Under $H_1$, the sum over the block has mean $\binom k2\mu$ and variance $\binom k2\Var_P(Y^{(l)}_{ij})$, and by the central limit theorem these two numbers determine its distribution when $k$ is large. The spectral test uses the largest eigenvalue of the $n\times n$ score matrix $\mathbf Y$, which we can write as $\mathbf Y=\E[\mathbf Y]+(\mathbf Y-\E[\mathbf Y])$. The expected matrix $\E[\mathbf Y]$ has the entry $\mu$ for every pair in the block and $0$ elsewhere, and its largest eigenvalue is $\mu(k-1)$. The entries of $\mathbf Y-\E[\mathbf Y]$ outside the block have mean $0$ and variance $1$, and the largest eigenvalue of such a matrix is close to $2\sqrt n$ \citep{bai1988necessary}. When $\mu(k-1)>\sqrt n$, the largest eigenvalue of $\mathbf Y$ is close to $\mu(k-1) + n/(\mu(k-1))$, which exceeds the edge of the spectrum \citep{feral2007largest}. The distribution $P$ only affects the $\binom k2$ block entries of $\mathbf Y-\E[\mathbf Y]$. If $P$ has a finite fourth moment, the derivation is similar to Appendix~\ref{app:spectral}, and, when $k/n\to0$, the largest eigenvalue of $\mathbf Y/\sqrt n$ has the same limits as in Appendix~\ref{app:spectral}.

The information-theoretic boundary follows from the second moment of the likelihood ratio under independence of $p$-values, and depends on $P$ through $\chi^2(P \Vert Q)=\int_0^1 f^2_P-1$, where $f_P$ is the density of $P$. The computational boundary assumes the planted clique hypothesis and requires conditions on $P$, so it depends on $P$ through the Kullback--Leibler divergence. Two distributions with the same $\mu$ can have different $\chi^2(P \Vert Q)$, and $\mu$ gives only the bound $\sqrt{\chi^2(P \Vert Q)}\ge\mu$. Hence a layer with $\hat\mu\ge\mu_{\mathrm{comp}}$ is easy for any $P$. A layer with $\hat\mu$ below the information-theoretic boundary or below $\mu_{\mathrm{comp}}$ is impossible or hard only if $\sqrt{\chi^2(P \Vert Q)}$ and $\sqrt{\dskl}$ are close to $\mu$; otherwise a test that uses $P$ can detect the block.

\textbf{Several modalities.} A group that coordinates in $M$ modalities forms a block in each of $M$ layers. Assume the simplest case where the $M$ layers carry the same signal $\mu$, their scores are independent, and every pair is tested in all $M$ layers. Then, the \emph{summed score} $\sum_{l=1}^MY^{(l)}_{ij}$ has null mean $0$ and null variance $M$, and its block entries have mean $M\mu$. By dividing by $\sqrt M$, the summed score has null variance $1$ and the mean of the block becomes $\sqrt M\mu$. This division does not change the decision of any test. The summed matrix therefore behaves as one layer with signal $\sqrt M\mu$, and every boundary is multiplied by $1/\sqrt M$: a coordinating group whose signal is below the information-theoretic boundary in each modality can be detectable when the signal is aggregated over many modalities. The null distribution of the summed score is a shifted Gamma distribution with shape parameter $M$, so the factor $\sqrt M$ is exact for the sum and spectral boundaries and holds to leading order for the scan boundary and the information-theoretic boundary. A simulation of the scan with $n=30$, $k\in\{3,4,5\}$, and $M\le256$ is consistent with this result: the signal at which the scan has power $0.5$ at level $0.05$ has a fitted slope against $M$ on a log--log scale between $-0.54$ and $-0.52$ for $M\ge16$, closest to the limit $-1/2$ for the largest $k$. 

\subsection{Signal Estimation and Group Detection}
\label{sec:predict}

The \emph{layer graph} of an operation and modality under one null model has as nodes the $n_l$ accounts that co-act with at least one other account. Let $k$ be the number of labelled accounts of the operation, and $S_l$ be the set of $k_l$ labelled accounts in the layer; $k_l\le k$, since a member that does not co-act in the modality is not in the layer. We estimate the signal strength $\mu$ of the block by the mean score of the labelled pairs of the layer, treating $S_l$ as the planted set $\Sstar$,
\begin{equation}\label{eq:muhat}
\hat\mu=\binom{k_l}{2}^{-1}\sum_{i<j\in S_l}Y^{(l)}_{ij},
\end{equation}
where an untested pair contributes $0$. A layer with $\hat\mu\le0$ has no measurable signal, because its labelled pairs score no higher on average than independent pairs; we report it as \emph{no signal}. The \emph{control mean} of a layer is computed like $\hat\mu$ over the pairs of control accounts, and is $0$ under a correctly specified null model.

We cap the $Y^{(l)}_{ij}$ scores at $20$, because the far tail of an estimated null model is not validated and a few extreme pairs would otherwise determine $\hat\mu$. The cap is above the largest score expected among null pairs: the largest layer has $9.0\times10^7$ tested pairs, and the expected number of null pairs above the cap is $9.0\times10^7e^{-21}<0.07$. One shared item gives a score of at most $\log(B+1)-1=3.62$, so a pair reaches the cap only if it shares at least nine items. The cap affects between $0.004\%$ to $1.6\%$ of the tested pairs of a layer.

\textbf{Recovery heuristics.} The scan over all $\binom{n}{k}$ sets is computationally infeasible for most data sets, so we define two greedy heuristics. \emph{Size-agnostic peeling} repeatedly removes the node whose removal most increases the mean score of the remaining set, and returns the set of at least $5$ accounts that maximises the density $\rho(S) = \sum_{i < j \in S} Y^{(l)}_{ij}/|S|$. The \emph{fixed-size heuristic} is given the size $k_l \leq k$ of the block in the layer, seeds a set of $k_l$ nodes from the peeling order, and improves $\rho(S)$ by single swaps (at most $200$ iterations), keeping the best of $3$ restarts. A permutation test compares the density $\rho$ of the returned set with the densities of the sets that the same heuristic returns after the edge weights are shuffled ($100$ replicates). The returned set $S$ \emph{beats chance} if a random set of $|S|$ nodes contains at least as many labelled nodes with probability below $0.05$ (hypergeometric test). We also report the $F_1$ score of $S$ against the labelled nodes, and the signal $\hat\mu(S)$ of the returned set, computed as in Equation~\eqref{eq:muhat} with $S$ in place of $S_l$. For the \textit{spectral method}, we compute the eigenvector of the largest eigenvalue $\nu_1$ of the score matrix of the layer, choose its sign as the sign whose $k_l$ largest entries have the larger total score, and return the $k_l$ accounts with the largest entries. A permutation test compares $\nu_1$ with its values after the edge weights are shuffled ($100$ replicates).

\section{Results}
\label{sec:results}

\begin{table}[ht]
\centering
\caption{Recovery by predicted regime: layers in which the returned set beats chance, median $F_1$ against the $k_l$ labelled accounts in the layer, and layers in which the permutation test rejects ($p \leq 0.05$; for the spectral method, the test on $\nu_1$). agn.: size-agnostic; fix.: fixed-size; spec.: spectral.}
\label{tab:recovery}
\footnotesize\setlength{\tabcolsep}{3pt}
\resizebox{\columnwidth}{!}{%
\begin{tabular}{lrccccccccc}
 & & \multicolumn{3}{c}{beats chance} & \multicolumn{3}{c}{median $F_1$} & \multicolumn{2}{c}{test rejects}\\
\cmidrule(lr){3-5}\cmidrule(lr){6-8}\cmidrule(lr){9-10}
regime & layers & agn. & fix. & spec. & agn. & fix. & spec. & agn. & spec.\\
\midrule
\multicolumn{10}{l}{\emph{uniform-modality null model}}\\
impossible & 6 & $0$ & $0$ & $0$ & $0.00$ & $0.00$ & $0.00$ & $6$ & $6$\\
hard & 25 & $4$ & $6$ & $10$ & $0.00$ & $0.00$ & $0.03$ & $22$ & $24$\\
easy & 168 & $123$ & $126$ & $123$ & $0.34$ & $0.59$ & $0.62$ & $153$ & $156$\\
no signal & 0 & -- & -- & -- & -- & -- & -- & -- & --\\
\midrule
\multicolumn{10}{l}{\emph{uniform-account null model}}\\
impossible & 10 & $1$ & $3$ & $0$ & $0.00$ & $0.00$ & $0.00$ & $10$ & $10$\\
hard & 26 & $10$ & $11$ & $12$ & $0.01$ & $0.03$ & $0.02$ & $24$ & $25$\\
easy & 162 & $132$ & $139$ & $134$ & $0.35$ & $0.67$ & $0.66$ & $157$ & $156$\\
no signal & 1 & $0$ & $0$ & $0$ & $0.01$ & $0.01$ & $0.00$ & $1$ & $1$\\
\midrule
\multicolumn{10}{l}{\emph{account-activity null model}}\\
impossible & 51 & $15$ & $15$ & $18$ & $0.00$ & $0.00$ & $0.00$ & $47$ & $46$\\
hard & 24 & $8$ & $7$ & $7$ & $0.01$ & $0.00$ & $0.00$ & $21$ & $21$\\
easy & 107 & $82$ & $89$ & $76$ & $0.26$ & $0.59$ & $0.56$ & $105$ & $105$\\
no signal & 17 & $7$ & $10$ & $10$ & $0.08$ & $0.22$ & $0.17$ & $17$ & $16$\\
\midrule
\multicolumn{10}{l}{\emph{item-virality null model}}\\
impossible & 14 & $3$ & $2$ & $4$ & $0.00$ & $0.00$ & $0.00$ & $14$ & $14$\\
hard & 13 & $6$ & $4$ & $8$ & $0.01$ & $0.00$ & $0.04$ & $13$ & $13$\\
easy & 72 & $42$ & $38$ & $39$ & $0.17$ & $0.28$ & $0.35$ & $72$ & $72$\\
no signal & 100 & $33$ & $27$ & $40$ & $0.03$ & $0.12$ & $0.23$ & $85$ & $87$\\
\bottomrule
\end{tabular}}
\end{table}

\begin{figure*}[ht]
\centering
\includegraphics[width=\textwidth]{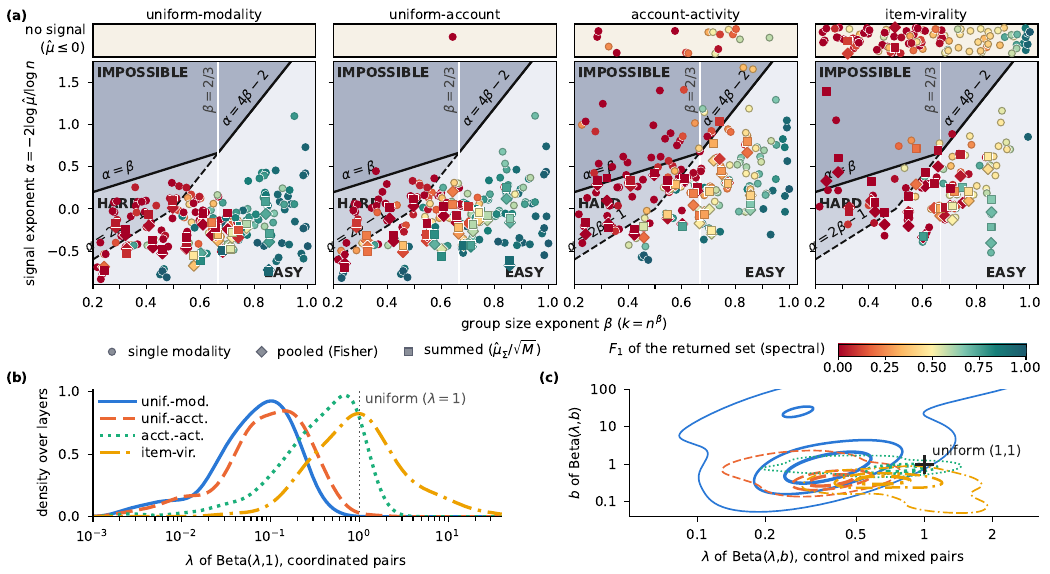}
\caption{\textbf{(a)} Phase diagrams under the four null models (uniform-modality, uniform-account, account-activity, item-virality). Each point is a graph layer, coloured by the $F_1$ of the set returned by the spectral method. \textbf{(b, c)} Beta distributions fitted by maximum likelihood to the $p$-values under each null model. \textbf{(b)} Density of $\lambda$ estimates for $\mathrm{Beta}(\lambda,1)$ fitted to the pairs labelled as coordinated. \textbf{(c)} Density of $(\lambda,b)$ estimates of $\mathrm{Beta}(\lambda,b)$ fitted to $p$-values between control pairs or mixed pairs, as contours at $5\%$, $35\%$, and $70\%$ of its maximum; the $+$ marker shows the uniform distribution, expected under a calibrated null model.
}
\label{fig:phasemap}
\end{figure*}

\begin{table*}[ht]
\centering
\caption{Control mean $\hat\mu$ (median and range over the operations) and coordination $\hat\mu$ of the labelled pairs (mean and range over the operations) by modality under each null model. The control mean is $0$ under a correctly specified null model.}
\label{tab:nullcompare}
\scriptsize\setlength{\tabcolsep}{2pt}
\resizebox{\textwidth}{!}{%
\begin{tabular}{lllllllll}
 & \multicolumn{4}{c}{control mean $\hat\mu$, median [range]} & \multicolumn{4}{c}{coordination $\hat\mu$, mean [range]}\\
\cmidrule(lr){2-5}\cmidrule(lr){6-9}
layer & account & item & unif.\ modality & unif.\ account & account & item & unif.\ modality & unif.\ account\\
\midrule
pooled & $0.012$ [$0.005$, $0.076$] & $0.006$ [$-0.031$, $0.16$] & $0.215$ [$0.061$, $1.48$] & $0.114$ [$0.040$, $0.56$] & $1.13$ [$0.046$, $3.56$] & $1.10$ [$-0.41$, $5.67$] & $5.78$ [$0.93$, $15.09$] & $4.04$ [$0.48$, $11.78$]\\
mentions & $0.001$ [$-0.003$, $0.014$] & $-0.001$ [$-0.022$, $0.12$] & $0.098$ [$0.010$, $1.27$] & $0.032$ [$0.003$, $0.44$] & $0.56$ [$-0.047$, $7.34$] & $0.54$ [$-0.53$, $4.09$] & $4.04$ [$0.53$, $15.03$] & $2.71$ [$0.25$, $14.75$]\\
hashtags & $0.011$ [$0.004$, $0.037$] & $0.001$ [$-0.029$, $0.17$] & $0.167$ [$0.041$, $0.50$] & $0.075$ [$0.018$, $0.27$] & $0.65$ [$-0.012$, $3.08$] & $1.23$ [$-0.55$, $7.57$] & $4.79$ [$0.52$, $14.38$] & $3.34$ [$-0.15$, $11.10$]\\
reply (acct) & $0.000$ [$-0.005$, $0.004$] & $-0.002$ [$-0.027$, $-0.0001$] & $0.008$ [$0.003$, $0.089$] & $0.002$ [$-0.018$, $0.025$] & $0.062$ [$-0.075$, $0.55$] & $-0.012$ [$-0.56$, $0.56$] & $1.64$ [$0.22$, $6.02$] & $0.99$ [$0.091$, $2.95$]\\
reply (post) & $0.016$ [$0.004$, $0.28$] & $-0.009$ [$-0.14$, $-0.001$] & $0.058$ [$0.012$, $0.52$] & $0.024$ [$0.006$, $0.41$] & $0.44$ [$0.009$, $2.30$] & $-0.095$ [$-0.24$, $-0.004$] & $1.17$ [$0.019$, $3.80$] & $1.07$ [$0.019$, $3.72$]\\
repost (acct) & $0.001$ [$-0.006$, $0.014$] & $-0.001$ [$-0.029$, $0.046$] & $0.052$ [$0.006$, $0.76$] & $0.018$ [$0.002$, $0.25$] & $0.48$ [$-0.098$, $6.56$] & $0.29$ [$-0.42$, $2.95$] & $3.43$ [$0.17$, $13.83$] & $2.41$ [$0.091$, $12.43$]\\
repost (post) & $0.005$ [$0.0005$, $0.062$] & $-0.002$ [$-0.052$, $-0.0002$] & $0.015$ [$0.002$, $0.18$] & $0.009$ [$0.0009$, $0.12$] & $1.61$ [$0.043$, $7.38$] & $-0.15$ [$-0.91$, $0.19$] & $3.55$ [$0.16$, $14.61$] & $3.43$ [$0.15$, $14.60$]\\
URLs & $0.005$ [$0.0004$, $0.13$] & $-0.002$ [$-0.14$, $-0.0003$] & $0.021$ [$0.002$, $0.52$] & $0.010$ [$0.0009$, $0.23$] & $0.85$ [$0.032$, $3.63$] & $-0.072$ [$-0.63$, $0.47$] & $2.43$ [$0.18$, $9.77$] & $2.17$ [$0.16$, $9.49$]\\
\midrule
all layers & $0.005$ [$-0.006$, $0.28$] & $-0.002$ [$-0.14$, $0.17$] & $0.055$ [$0.002$, $1.48$] & $0.021$ [$-0.018$, $0.56$] & $0.73$ [$-0.098$, $7.38$] & $0.37$ [$-0.91$, $7.57$] & $3.43$ [$0.019$, $15.09$] & $2.57$ [$-0.15$, $14.75$]\\
\end{tabular}}
\end{table*}

The predicted regime of the 26 labelled information operations depends strongly on the chosen null model (Figure~\ref{fig:phasemap}, Table~\ref{tab:nullcompare}). 
The item-virality null model places the fewest layers in the easy regime ($36\%$), 
followed by the account-activity null model ($54\%$),
 the uniform-account null model ($81\%$)
and the uniform-modality null model ($84\%$; Table~\ref{tab:recovery}). These differences stem from the fact that the null models differ in which timing patterns they attribute to chance. The item-virality null model redraws the action times from the adoption curve of the item. Because the data contain only a sample of control accounts, many of the post and URL items have exactly two adopters, and on such an item the item-virality null model gives $p=1$ when each account acts once. As a result, this null model measures no signal in 
$50\%$ of layers, and the $p$-values of the labelled pairs follow $\mathrm{Beta}(\lambda,1)$ with a median $\lambda$ of $0.98$, close to the uniform distribution. The account-activity null model keeps the hourly profile of each account, but it does not reproduce the independent reactions of different accounts to a new popular item, so the control mean is slightly positive in all $71$ reply (post), repost (post), and URL layers. The uniform null models keep only the period in which a modality or an account is active. They measure the strongest signal for the information operations
(see Table~\ref{tab:nullcompare} and Figure~\ref{fig:phasemap}b).
However, their control mean is positive in almost all layers, and the $p$-values of the control pairs are further from uniform than under the account-activity null model (see Figure~\ref{fig:phasemap}c, median $\lambda$ of $0.37$ and $0.36$, against $0.74$).
That is, the choice of a null model is a compromise between distinguishing the coordinated group and the control group: a null model that attributes more timing patterns to chance follows better the uniform $p$-value distribution for control accounts, but also absorbs the coordination itself, while one that attributes fewer patterns leaves organic synchrony in the scores but separates the coordinated pairs more strongly.

\begin{figure}[ht]
\centering
\includegraphics[width=\columnwidth]{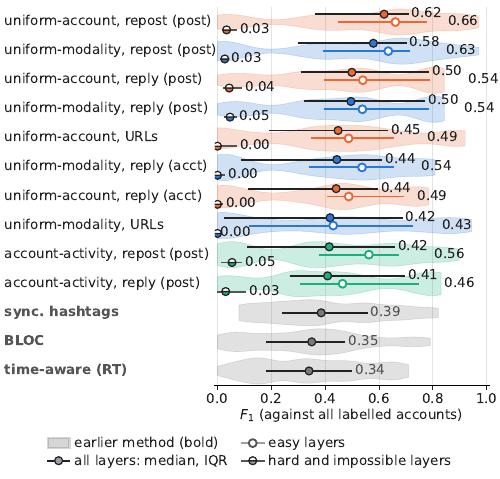}
\caption{$F_1$ of the spectral method for each null model and layer that scores higher than the top 3 methods in \citep{iannucci2026detecting}.}
\label{fig:bench}
\end{figure}

Recovery of the ground-truth groups using the spectral method exceeds that of all $20$ methods that \citet{iannucci2026detecting} evaluate on the same $26$ operations (Figure~\ref{fig:bench}). As in their evaluation, we compute $F_1$ against all labelled accounts of the operation. Both evaluations use the ground truth beyond the final score: the spectral method is given the size $k_l$ of the labelled group in the layer, and $F_1^*$ \citep{iannucci2026detecting} is the $F_1$ of the cluster that best matches the labelled accounts. When each layer and null model is considered separately, the spectral method exceeds all $20$ methods in $10$ of the $28$ single-modality combinations. The highest median $F_1$ is $0.62$ for reposts of a post under the uniform-account null model, compared to $0.39$ for synchronised hashtags \citep{magelinski2022synchronized} and $0.34$ for time-aware (RT) \citep{iannucci2026detecting}, which uses the same modality.
Some of the layers carry poor signal over the operations, but even when evaluated over all operations and single-modality layers, the spectral method reaches a median $F_1$ of $0.40$ under the uniform-account null model, against $0.36$, $0.24$, and $0.18$ under the uniform-modality, account-activity, and item-virality null models, and against $0.39$, the highest reported median $F_1$ \citep{iannucci2026detecting}, achieved by synchronised hashtags. 
On the layers that the theory places in the easy regime, the spectral method exceeds all of the previous methods \citep{iannucci2026detecting} under every null model, with a median $F_1$ of $0.45$, $0.48$, $0.46$, and $0.39$ under the uniform-modality, uniform-account, account-activity, and item-virality null models, and a median recall of $0.34$ to $0.38$; on the hard and impossible layers, the median $F_1$ is at most $0.04$. 

Under all four null models, combining modalities moves operations toward the easy regime, but with less than the factor $1/\sqrt M$ derived in Section~\ref{sec:threeregimes}. We combine the modalities in two ways. The \emph{summed layer} adds the scores of a pair over the $M=7$ modalities, and is placed on the phase diagram at $\hat\mu_\Sigma/\sqrt M$. The \emph{pooled layer} combines the $p$-values of a pair across all modalities by Fisher's method, so, under our assumptions, every tested pair has expectation $0$ and variance $1$ under $H_0$, independently of the number of layers. Since Fisher's statistic $-2 \sum \log p$ equals, up to a constant, the sum of the scores, the two layers differ only in how the sum is standardised. Under the account-activity, item-virality, uniform-modality, and uniform-account null models, the pooled layer is easy in $18$, $18$, $23$, and $22$ of the $26$ operations, and the summed layer in $17$, $14$, $22$, and $22$, in both cases more often than the single-modality layers. The regime of the pooled layer also depends less on the null model: the four null models agree on $65\%$ of the
operations, against 
$22\%$ of the
single-modality layers. However, the boundaries move less than the factor $1/\sqrt M$ for two reasons. First, our derivation requires every pair to be tested in all $M$ layers, while a pair is tested in a mean of $1.34$ layers (median over the operations; range $1.04$ to $2.0$). Second, our derivation assumes one block with the same signal in every layer, while, in practice, we observe that each modality involves a different subset of the operation. Thus, a combined layer averages the signal of the strongest modality with that of the others. Under no null model is an operation easy on the pooled layer without being easy in some single modality. Recovery is lower on the combined layers: the median $F_1$ of the fixed-size heuristic is $0.03$, $0.19$, $0.12$, and $0.02$ on the pooled layer and $0.18$, $0.39$, $0.29$, and $0.02$ on the summed layer, against $0.60$, $0.69$, $0.64$, and $0.37$ on the easy single-modality layers (uniform-modality, uniform-account, account-activity, and item-virality null models), although a combined layer is scored against nearly the whole operation in a larger population.

Under all four null models, the heuristics find a set denser than chance in almost every layer, whether or not this set is the operation (Table~\ref{tab:recovery}). Even on layers where the found set does not beat chance recovery, the found set is still denser than chance in the vast majority of cases. The sets found by the size-agnostic heuristic, fixed-size heuristic, and spectral method are also denser than the operation in most layers. The pooled layer falls under the easy regime in most operations under every null model, but the spectral set has a median precision of $0.06$, $0.22$, $0.10$, and $0.04$, and a signal $1.9$, $1.5$, $1.5$, and $3.8$ times that of the labelled pairs. The scores therefore contain dense groups of mostly control accounts, which the model of one block in independent scores excludes. A significant permutation test only shows that a layer contains a dense set, but does not guarantee that the dense set represents the information operation.

\section{Conclusion}
\label{sec:conclusion}

We treat coordination detection as hypothesis testing across all accounts, reducing the task to detecting a planted block within a matrix of scores. Our derived boundaries establish the signal thresholds below which detection, and therefore recovery, is impossible, and above which polynomial-time tests detect the block. Our derivation shows how combining scores across multiple modalities lowers the statistical and computational boundaries. We demonstrate that algorithmic recovery is bounded by the chosen null model, through the timing patterns that it attributes to chance and the signal that it can measure. Across 26 labelled information operations, null models that attribute fewer patterns to chance are less calibrated, but they separate coordinated from organic pairs more strongly. Consequently, under these models, a spectral method recovers the operations better, by median $F_1$, than previously proposed coordination detection methods (Figure \ref{fig:bench}).

The dependence on the null model reveals two criticalities that affect the current study on coordination detection. First, permutation tests find a significantly dense set across nearly all layers, so the existence of a dense cluster in a coordination network is not evidence of coordination. Second, the simplest null models, that keep only the active period and attribute every timing pattern to synchrony, recover the coordinated groups best. Such null models separate the labelled accounts from the control accounts only based on their timing, which points to a sampling artefact in benchmark datasets: the accounts of these operations are not hiding in the timing patterns of organic accounts. This can reflect detection bias of the platform itself, so that information operations accounts that hide better are absent from the ground truth, or the collection of the data. The control accounts were sampled by the hashtags and dates of each operation, with at most $100$ posts on those dates \citep{seckin2025labeled}, and the uniform-account null model spreads the actions of an account over its own active period, so the same co-action is far more extreme for two accounts of the operation than for two control accounts.
This calls for additional benchmark datasets, so that coordination detection frameworks can be evaluated against realistic conditions and more varied strategies and patterns of coordination.

\textbf{Limitations.} Our approach focuses on temporal synchronisation conditional on accounts interacting with the same items, rather than coordination expressed through unusual co-adoption. The null models account separately for individual activity patterns, item virality, or only the active period, but none controls for both activity patterns and item virality simultaneously. A joint null model could address this limitation, although it would require additional modelling assumptions and risk absorbing genuine coordination into the estimated baseline. Conversely, considering each null model separately allows us to examine how different sources of chance synchronisation affect the conclusions. Exploring joint null models through stratified or weighted randomisation is a direction for future work. 

Our theoretical formulation and empirical evaluation rely on simplifying assumptions. First, on a statistical level, we assume independent $p$-values, whereas pairs that share an account are correlated. Furthermore, because Monte Carlo $p$-values are discrete and only approximately uniform under the null model, they yield slight biases and cause score means and variances to be $0$ and $1$ only approximately. However, because real-world detection lacks labels, the deviation from calibration cannot simply be solved by empirical null distributions. Instead, this calls for further research on null models that accurately capture organic timing without absorbing the coordination signal itself. Second, our derivation relies on the planted submatrix problem, assuming a single, homogeneous block of coordinated accounts within a fully tested matrix. Because real-world information operations may be complex and consist of multiple sub-groups, future research could extend the derivation of phase boundaries to nested block structures. Similarly, the multi-modal gain assumes a uniform group producing equal signal in all modalities. Deriving the signal gain for coordination that alternates across modalities, or modalities with unequal signal, is left for future work. Finally, since the greedy heuristics and the spectral method are empirical approximations of the exhaustive scan and polynomial-time tests used in the theoretical derivation, their failure in the easy regime reflects either algorithmic limitations or, where $\mu_{\mathrm{sum}}\le\hat\mu<\mu_{\mathrm{spec}}$, the gap between detection and recovery (Section \ref{sec:threeregimes}). Since the densest subgraph problem with negative edge weights is NP-hard \citep{tsourakakis2019novel}, and our scores take negative values, good heuristics for weighted signed multiplex networks are essential to bridge theoretical detectability and empirical detection.

\bibliographystyle{unsrtnat}
\bibliography{references}

@article{fortunato2010community,
  title={Community detection in graphs},
  author={Fortunato, Santo and others},
  year={2010},
  publisher={Elsevier}
}

@article{ferrara2016rise,
  title={The rise of social bots},
  author={Ferrara, Emilio and Varol, Onur and Davis, Clayton and Menczer, Filippo and Flammini, Alessandro},
  journal={Communications of the ACM},
  volume={59},
  number={7},
  pages={96--104},
  year={2016},
  publisher={ACM New York, NY, USA}
}

@article{starbird2019disinformation,
  title={Disinformation as collaborative work: Surfacing the participatory nature of strategic information operations},
  author={Starbird, Kate and Arif, Ahmer and Wilson, Tom},
  journal={Proceedings of the ACM on Human-Computer Interaction},
  volume={3},
  number={CSCW},
  pages={1--26},
  year={2019},
  publisher={ACM New York, NY, USA}
}

@article{keller2020political,
  title={Political astroturfing on twitter: How to coordinate a disinformation campaign},
  author={Keller, Franziska B and Schoch, David and Stier, Sebastian and Yang, JungHwan},
  journal={Political communication},
  volume={37},
  number={2},
  pages={256--280},
  year={2020},
  publisher={Taylor \& Francis}
}

@inproceedings{pacheco2021uncovering,
  title={Uncovering coordinated networks on social media: methods and case studies},
  author={Pacheco, Diogo and Hui, Pik-Mai and Torres-Lugo, Christopher and Truong, Bao Tran and Flammini, Alessandro and Menczer, Filippo},
  booktitle={Proceedings of the international AAAI conference on web and social media},
  volume={15},
  pages={455--466},
  year={2021}
}

@article{magelinski2022synchronized,
  title={A synchronized action framework for detection of coordination on social media},
  author={Magelinski, Thomas and Ng, Lynnette and Carley, Kathleen},
  journal={Journal of Online Trust and Safety},
  volume={1},
  number={2},
  year={2022}
}

@inproceedings{sharma2021identifying,
  title={Identifying coordinated accounts on social media through hidden influence and group behaviours},
  author={Sharma, Karishma and Zhang, Yizhou and Ferrara, Emilio and Liu, Yan},
  booktitle={Proceedings of the 27th ACM SIGKDD conference on knowledge discovery \& data mining},
  pages={1441--1451},
  year={2021}
}

@article{tardelli2024multifaceted,
  title={Multifaceted online coordinated behavior in the 2020 US presidential election},
  author={Tardelli, Serena and Nizzoli, Leonardo and Avvenuti, Marco and Cresci, Stefano and Tesconi, Maurizio},
  journal={EPJ Data Science},
  volume={13},
  number={1},
  pages={33},
  year={2024},
  publisher={Springer Berlin Heidelberg}
}

@article{cima2024coordinated,
  title={Coordinated behavior in information operations on Twitter},
  author={Cima, Lorenzo and Mannocci, Lorenzo and Avvenuti, Marco and Tesconi, Maurizio and Cresci, Stefano},
  journal={IEEE Access},
  volume={12},
  pages={61568--61585},
  year={2024},
  publisher={IEEE}
}

@article{mannocci2025multimodal,
  title={Multimodal Coordinated Online Behavior: Trade-offs and Strategies},
  author={Mannocci, Lorenzo and Cresci, Stefano and Magnani, Matteo and Monreale, Anna and Tesconi, Maurizio},
  journal={arXiv:2507.12108},
  year={2025}
}

@article{ng2022combined,
  title={A combined synchronization index for grassroots activism on social media},
  author={Ng, Lynnette Hui Xian and Carley, Kathleen M},
  journal={arXiv:2212.13221},
  year={2022}
}

@inproceedings{luceri2024unmasking,
  title={Unmasking the web of deceit: Uncovering coordinated activity to expose information operations on twitter},
  author={Luceri, Luca and Pant{\`e}, Valeria and Burghardt, Keith and Ferrara, Emilio},
  booktitle={Proceedings of the ACM Web Conference 2024},
  pages={2530--2541},
  year={2024}
}

@article{weber2021amplifying,
  title={Amplifying influence through coordinated behaviour in social networks},
  author={Weber, Derek and Neumann, Frank},
  journal={Social Network Analysis and Mining},
  volume={11},
  number={1},
  pages={111},
  year={2021},
  publisher={Springer}
}

@inproceedings{seckin2025labeled,
  title={Labeled datasets for research on information operations},
  author={Seckin, Ozgur Can and Pote, Manita and Nwala, Alexander C and Yin, Lake and Luceri, Luca and Flammini, Alessandro and Menczer, Filippo},
  booktitle={Proceedings of the International AAAI Conference on Web and Social Media},
  volume={19},
  pages={2567--2574},
  year={2025}
}

@inproceedings{hooi2016fraudar,
  title={Fraudar: Bounding graph fraud in the face of camouflage},
  author={Hooi, Bryan and Song, Hyun Ah and Beutel, Alex and Shah, Neil and Shin, Kijung and Faloutsos, Christos},
  booktitle={Proceedings of the 22nd ACM SIGKDD international conference on knowledge discovery and data mining},
  pages={895--904},
  year={2016}
}

@inproceedings{charikar2000greedy,
  title={Greedy approximation algorithms for finding dense components in a graph},
  author={Charikar, Moses},
  booktitle={International workshop on approximation algorithms for combinatorial optimization},
  pages={84--95},
  year={2000},
  organization={Springer}
}

@inproceedings{tsourakakis2019novel,
  title={Novel dense subgraph discovery primitives: Risk aversion and exclusion queries},
  author={Tsourakakis, Charalampos E and Chen, Tianyi and Kakimura, Naonori and Pachocki, Jakub},
  booktitle={Joint European Conference on Machine Learning and Knowledge Discovery in Databases},
  pages={378--394},
  year={2019},
  organization={Springer}
}

@inproceedings{brennan2020reducibility,
  title={Reducibility and statistical-computational gaps from secret leakage},
  author={Brennan, Matthew and Bresler, Guy},
  booktitle={Conference on Learning Theory},
  pages={648--847},
  year={2020},
  organization={PMLR}
}

@inproceedings{nizzoli2021coordinated,
  title={Coordinated behavior on social media in 2019 UK general election},
  author={Nizzoli, Leonardo and Tardelli, Serena and Avvenuti, Marco and Cresci, Stefano and Tesconi, Maurizio},
  booktitle={Proceedings of the international AAAI conference on web and social media},
  volume={15},
  pages={443--454},
  year={2021}
}

@inproceedings{iannucci2026detecting,
  title={Detecting Coordinated Activities Through Temporal, Multiplex, and Collaborative Analysis},
  author={Iannucci, Letizia and Muratore, Elisa and Matakos, Antonis and Kivel{\"a}, Mikko},
  booktitle={Proceedings of the International AAAI Conference on Web and Social Media},
  volume={20},
  number={1},
  pages={1063--1082},
  year={2026}
}

@article{arias2014community,
  title={Community detection in dense random networks},
  author={Arias-Castro, Ery and Verzelen, Nicolas},
  journal={The Annals of Statistics},
  pages={940--969},
  year={2014},
  publisher={JSTOR}
}

@article{arias2011detection,
  title={Detection of an anomalous cluster in a network},
  author={Arias-Castro, Ery and Candes, Emmanuel J and Durand, Arnaud},
  journal={The Annals of Statistics},
  pages={278--304},
  year={2011},
  publisher={JSTOR}
}

@article{butucea2013detection,
  title={Detection of a sparse submatrix of a high-dimensional noisy matrix},
  author={Butucea, Cristina and Ingster, Yuri I},
  journal={Bernoulli},
  pages={2652--2688},
  year={2013},
  publisher={JSTOR}
}

@article{gauvin2022randomized,
  title={Randomized reference models for temporal networks},
  author={Gauvin, Laetitia and G{\'e}nois, Mathieu and Karsai, M{\'a}rton and Kivel{\"a}, Mikko and Takaguchi, Taro and Valdano, Eugenio and Vestergaard, Christian L},
  journal={Siam Review},
  volume={64},
  number={4},
  pages={763--830},
  year={2022},
  publisher={SIAM}
}

@article{kulldorff1997spatial,
  title={A spatial scan statistic},
  author={Kulldorff, Martin},
  journal={Communications in Statistics-Theory and methods},
  volume={26},
  number={6},
  pages={1481--1496},
  year={1997},
  publisher={Taylor \& Francis}
}

@article{ma2015computational,
  title={Computational barriers in minimax submatrix detection},
  author={Ma, Zongming and Wu, Yihong},
  journal={The Annals of Statistics},
  pages={1089--1116},
  year={2015},
  publisher={JSTOR}
}

@inproceedings{cinus2025exposing,
  title={Exposing cross-platform coordinated inauthentic activity in the run-up to the 2024 us election},
  author={Cinus, Federico and Minici, Marco and Luceri, Luca and Ferrara, Emilio},
  booktitle={Proceedings of the ACM on Web Conference 2025},
  pages={541--559},
  year={2025}
}

@inproceedings{pante2025beyond,
  title={Beyond interaction patterns: Assessing claims of coordinated inter-state information operations on Twitter/X},
  author={Pant{\`e}, Valeria and Axelrod, David and Flammini, Alessandro and Menczer, Filippo and Ferrara, Emilio and Luceri, Luca},
  booktitle={Companion Proceedings of the ACM on Web Conference 2025},
  pages={1234--1238},
  year={2025}
}

@article{mungari2026temporal,
  title={Temporal signatures of coordination: detecting and anticipating influence operations on social media},
  author={Mungari, Simone and Ferrara, Emilio and Luceri, Luca},
  journal={EPJ Data Science},
  year={2026},
  publisher={Springer}
}

@article{panayiotou2026setting,
  title={Setting the clock: Evaluating temporal window parameters for coordinated behavior detection},
  author={Panayiotou, Georgios and Mannocci, Lorenzo and Tesconi, Maurizio},
  journal={arXiv preprint arXiv:2609.21959},
  year={2026}
}

@article{mannocci2026detection,
  title={Detection and characterization of coordinated online behavior: A survey},
  author={Mannocci, Lorenzo and Mazza, Michele and Monreale, Anna and Tesconi, Maurizio and Cresci, Stefano},
  journal={ACM Computing Surveys},
  volume={58},
  number={16},
  pages={1--38},
  year={2026},
  publisher={ACM New York, NY}
}

@article{cai2017computational,
  title={Computational and statistical boundaries for submatrix localization in a large noisy matrix},
  author={Cai, T Tony and Liang, Tengyuan and Rakhlin, Alexander},
  year={2017}
}

@inproceedings{brennan2019universality,
  title={Universality of computational lower bounds for submatrix detection},
  author={Brennan, Matthew and Bresler, Guy and Huleihel, Wasim},
  booktitle={Conference on Learning Theory},
  pages={417--468},
  year={2019},
  organization={PMLR}
}

@inproceedings{hajek2015computational,
  title={Computational lower bounds for community detection on random graphs},
  author={Hajek, Bruce and Wu, Yihong and Xu, Jiaming},
  booktitle={Conference on Learning Theory},
  pages={899--928},
  year={2015},
  organization={PMLR}
}

@article{feral2007largest,
  title={The largest eigenvalue of rank one deformation of large Wigner matrices},
  author={F{\'e}ral, Delphine and P{\'e}ch{\'e}, Sandrine},
  journal={Communications in mathematical physics},
  volume={272},
  number={1},
  pages={185--228},
  year={2007},
  publisher={Springer}
}

@article{addario2010combinatorial,
  title={On combinatorial testing problems},
  author={Addario-Berry, Louigi and Broutin, Nicolas and Devroye, Luc and Lugosi, G{\'a}bor},
  year={2010}
}

@article{shabalin2009finding,
  title={Finding large average submatrices in high dimensional data},
  author={Shabalin, Andrey A and Weigman, Victor J and Perou, Charles M and Nobel, Andrew B},
  journal={The Annals of Applied Statistics},
  pages={985--1012},
  year={2009},
  publisher={JSTOR}
}

@article{hajek2018submatrix,
  title={Submatrix localization via message passing},
  author={Hajek, Bruce and Wu, Yihong and Xu, Jiaming},
  journal={Journal of Machine Learning Research},
  volume={18},
  number={186},
  pages={1--52},
  year={2018}
}

@article{chen2016statistical,
  title={Statistical-computational tradeoffs in planted problems and submatrix localization with a growing number of clusters and submatrices},
  author={Chen, Yudong and Xu, Jiaming},
  journal={Journal of Machine Learning Research},
  volume={17},
  number={27},
  pages={1--57},
  year={2016}
}

@book{lehmann2005testing,
  title={Testing statistical hypotheses},
  author={Lehmann, Erich Leo and Romano, Joseph P},
  year={2005},
  publisher={Springer}
}

@article{knowles2013isotropic,
  title={The isotropic semicircle law and deformation of Wigner matrices},
  author={Knowles, Antti and Yin, Jun},
  journal={Communications on Pure and Applied Mathematics},
  volume={66},
  number={11},
  pages={1663--1749},
  year={2013},
  publisher={Wiley Online Library}
}

@article{bai1988necessary,
  title={Necessary and sufficient conditions for almost sure convergence of the largest eigenvalue of a Wigner matrix},
  author={Bai, Zhi-Dong and Yin, Yong-Qua},
  journal={The Annals of Probability},
  pages={1729--1741},
  year={1988},
  publisher={JSTOR}
}

@book{tsybakov2009introduction,
  title={Introduction to nonparametric estimation},
  author={Tsybakov, Alexandre B and Zaiats, Vladimir},
  volume={11},
  year={2009},
  publisher={Springer}
}

@article{sellke2001calibration,
  title={Calibration of $\rho$ values for testing precise null hypotheses},
  author={Sellke, Thomas and Bayarri, Mar{\'\i}a Jes{\'u}s and Berger, James O},
  journal={The American Statistician},
  volume={55},
  number={1},
  pages={62--71},
  year={2001},
  publisher={Taylor \& Francis}
}

@article{heard2018choosing,
  title={Choosing between methods of combining-values},
  author={Heard, Nicholas A and Rubin-Delanchy, Patrick},
  journal={Biometrika},
  volume={105},
  number={1},
  pages={239--246},
  year={2018},
  publisher={Oxford University Press}
}

@article{guimera2004modularity,
  title={Modularity from fluctuations in random graphs and complex networks},
  author={Guimera, Roger and Sales-Pardo, Marta and Amaral, Lu{\'\i}s A Nunes},
  journal={Physical Review E—Statistical, Nonlinear, and Soft Matter Physics},
  volume={70},
  number={2},
  pages={025101},
  year={2004},
  publisher={APS}
}

@book{peixoto2023descriptive,
  title={Descriptive vs. inferential community detection in networks: Pitfalls, myths and half-truths},
  author={Peixoto, Tiago P},
  year={2023},
  publisher={Cambridge University Press}
}

@book{ingster2012nonparametric,
  title={Nonparametric goodness-of-fit testing under Gaussian models},
  author={Ingster, Yuri and Suslina, Irina A},
  year={2012},
  publisher={Springer Science \& Business Media}
}

@article{montanari2015limitation,
  title={On the limitation of spectral methods: From the gaussian hidden clique problem to rank-one perturbations of gaussian tensors},
  author={Montanari, Andrea and Reichman, Daniel and Zeitouni, Ofer},
  journal={Advances in Neural Information Processing Systems},
  volume={28},
  year={2015}
}

@article{fulton2000eigenvalues,
  title={Eigenvalues, invariant factors, highest weights, and Schubert calculus},
  author={Fulton, William and others},
  journal={BULLETIN-AMERICAN MATHEMATICAL SOCIETY},
  volume={37},
  number={3},
  pages={209--250},
  year={2000},
  publisher={AMERICAN MATHEMATICAL SOCIETY}
}

\appendix

\section{Derivation of the boundaries}
\label{app:derivation}

\subsection{Assumptions}
\label{app:assumptions}

The derivations use the model of Section~\ref{sec:coord-as-planted} for one layer, under these assumptions:

\begin{itemize}
    \item[(A1)] the scores $Y_{ij}$, $i<j$, are independent;
    \item[(A2)] under $H_0$, every $p_{ij}$ has the uniform distribution $Q$;
    \item[(A3)] under $H_1$, $\Sstar$ is uniform on $\mathcal S_k$ with $p_{ij}\sim P = \mathrm{Beta}(\lambda,1)$ for $i,j\in\Sstar$, and $p_{ij} \sim Q$ otherwise;
    \item[(A4)] $k$ is known;
    \item[(A5)] every pair is tested. In the data, a pair that does not co-act has score $0$ under both hypotheses. The diagonal of $\mathbf Y$ is $0$ for the spectral test.
\end{itemize}

$\mathbb{P}_0$, $\mathbb{P}_1$ and $\E_0$, $\E_1$ denote the distributions of $Y$ and the expectations under $H_0$, $H_1$. Asymptotic statements refer to $n\to\infty$ with $k=n^\beta$ and $\mu^2=n^{-\alpha}$ for fixed $0<\beta<1$ and $\alpha$; a formula holds \emph{to leading order} if the ratio of its two sides tends to $1$.

\subsection{Distribution of the scores}
\label{app:score}

The log-likelihood ratio of $\mathrm{Beta}(\lambda,1)$ against the uniform distribution, evaluated at the observed $p^{(l)}_{ij}$, is

\begin{equation}
    \log(\ell(p^{(l)}_{ij})) = \log\lambda+(1-\lambda)(-\log p^{(l)}_{ij})\,.
\label{eq:llr}
\end{equation}

The likelihood ratio of a block entry against a null entry is $\ell=dP/dQ$. As a function of the score, $\ell(Y_{ij})=\lambda e^{(1-\lambda)(Y_{ij}+1)}$, the exponential of Equation~\eqref{eq:llr}. 

Since the score is a one-to-one function of the $p$-value, $\ell$ takes the same value whether it is computed from $p_{ij}$ or from $Y_{ij}$. Every divergence between $P$ and $Q$ and every condition in \citep{brennan2019universality} depends on the two distributions only through $\ell$, and therefore has the same value for the $p$-values and for the scores. 

For a set $S$ with $m=\binom k2$ pairs and $\Sigma(S)=\sum_{i<j\in S}Y_{ij}$: 
\begin{align}
H_0:&\ \Sigma(S)+m\sim\mathrm{Gamma}(m,1),\notag\\
H_1:&\ \Sigma(\Sstar)+m\sim\mathrm{Gamma}(m,\lambda),
\label{eq:oracle-h1-exact}
\end{align}
with mean $0$ and variance $m$ under $H_0$, and mean $m\mu$ and variance $m(1+\mu)^2$ under $H_1$.

\subsection{Scan boundary}
\label{app:scanthresh}

Following the scan-statistic analysis of
\cite{arias2011detection,butucea2013detection}, we use a Chernoff bound and a union bound to derive the sufficient detection condition for the exhaustive scan stated in Section~\ref{sec:threeregimes}, adapting the calculation to symmetric matrices with shifted-exponential entries. 

A set $S\in\mathcal S_k$ contains $m=\binom k2$ pairs, and there are $|\mathcal S_k|=\binom nk$ such sets. For a set $S$, let $\Sigma(S)=\sum_{i<j\in S}Y_{ij}$, so that $T_{\mathrm{scan}}=\max_{S\in\mathcal S_k}\Sigma(S)$, and let $L=\log(n/k)$. As $n\to\infty$ with $k=n^\beta$, $0<\beta<1$, and bounded $\mu$, we have $L\to\infty$ and $L/k\to0$.

Under $H_0$, $Y_{ij}+1\sim\mathrm{Exp}(1)$, so for $0\le t<1$ the logarithm of the moment generating function of a score is
\[
\psi_0(t)=\log\E_0\big[e^{tY_{ij}}\big]=-t-\log(1-t)=\sum_{r\ge2}\frac{t^r}{r}.
\]
Since $t^r/r\le t^r/2$ for $r\ge2$,

\begin{equation}
\psi_0(t)\le\frac{t^2}{2(1-t)},\qquad 0\le t<1.
\label{eq:psi-bound}
\end{equation}

For a fixed set $S$, $c>0$ and $0<t<1$, Markov's inequality applied to $e^{t\Sigma(S)}$ and the independence of the $m$ scores of $S$ give $\mathbb P_0(\Sigma(S)>c)\le\exp\big(-tc+m\,\psi_0(t)\big)$. The scan exceeds $c$ only if at least one set does, so a union bound over $\mathcal S_k$ and $\log\binom nk\le kL+k$ give

\begin{equation}
\begin{split}
\mathbb P_0\big(T_{\mathrm{scan}}>c\big)\le\binom nk\exp\big(-tc+m\,\psi_0(t)\big)\\
\le\exp\big(kL+k-tc+m\,\psi_0(t)\big),
\label{eq:chernoff}
\end{split}
\end{equation}
with $\psi_0(t)=\log\E_0[e^{tY_{ij}}]=-t-\log(1-t)=t^2/2+O(t^3)$. 

Let
\begin{equation}
c^\star=\sqrt{2mkL}=k\sqrt{(k-1)L},
\label{eq:wstar}
\end{equation}
and take $c=(1+\varepsilon)\,c^\star$ for a fixed $\varepsilon>0$, and $t=c/m=2(1+\varepsilon)\sqrt{L/(k-1)}$, which tends to $0$ because $L/k\to0$. By Equation~\eqref{eq:psi-bound}, $m\,\psi_0(t)\le c^2/\big(2m(1-t)\big)$, and with $tc=c^2/m$ and $c^2/(2m)=(1+\varepsilon)^2kL$,

\[
-tc+m\,\psi_0(t)\le-\frac{c^2}{2m}\cdot\frac{1-2t}{1-t}=-(1+\varepsilon)^2\,kL\,\frac{1-2t}{1-t}.
\]
Inserted into Equation~\eqref{eq:chernoff}, this gives
\[
\log\mathbb P_0\big(T_{\mathrm{scan}}>c\big)\le-kL\Big[(1+\varepsilon)^2\,\frac{1-2t}{1-t}-1-\frac1L\Big].
\]
As $n\to\infty$, the bracket tends to $(1+\varepsilon)^2-1>0$ and $kL\to\infty$, so $\mathbb P_0(T_{\mathrm{scan}}>c)\to0$.

Under $H_1$, the $m$ scores of $\Sstar$ are independent with $Y_{ij}+1\sim\mathrm{Exp}(\lambda)$, $\lambda=1/(1+\mu)$, so $\E_1[\Sigma(\Sstar)]=m\mu$ and $\Var_1(\Sigma(\Sstar))=m(1+\mu)^2$. The maximum over all sets is at least the sum over the block, $T_{\mathrm{scan}}\ge\Sigma(\Sstar)$. If $m\mu\ge(1+2\varepsilon)\,c^\star$, the event $\{T_{\mathrm{scan}}\le c\}$ implies $\Sigma(\Sstar)-m\mu\le-\varepsilon c^\star$, and Chebyshev's inequality gives

\[
\mathbb P_1\big(T_{\mathrm{scan}}\le c\big)\le\frac{m(1+\mu)^2}{(\varepsilon c^\star)^2}=\frac{(1+\mu)^2}{2\varepsilon^2kL}\;\longrightarrow\;0.
\]

The condition $m\mu\ge(1+2\varepsilon)\,c^\star$ reads $\mu\ge(1+2\varepsilon)\,\mu_{\mathrm{scan}}$, with
\begin{equation}
\mu_{\mathrm{scan}}=\frac{c^\star}{m}=2\sqrt{\frac{L}{k-1}}=2\sqrt{\frac{\log(n/k)}{k-1}}.
\label{eq:muach}
\end{equation}
Since $\varepsilon>0$ is arbitrary, for every fixed $\delta>0$ the scan with critical value $(1+\delta/2)\,c^\star$ has both error probabilities tending to $0$ when $\mu\ge(1+\delta)\,\mu_{\mathrm{scan}}$. This is a sufficient condition: the union bound treats overlapping sets as disjoint, so the exact boundary of the scan may be lower.

\subsection{Information-theoretic boundary}
\label{app:it-bound}

Following the second-moment approach of
\citep{arias2014community,butucea2013detection} and the testing inequalities of \citep{tsybakov2009introduction}, we derive the lower bound on total testing error underlying the information-theoretic bound in Section~\ref{sec:threeregimes} and examine its dependence on group size and signal strength.

\emph{Error bound.} A test decides from the score matrix $\mathbf Y$ whether to reject $H_0$. Let $A$ be the set of score matrices for which it rejects. The test makes a false positive with probability $\mathbb P_0(\mathbf Y\in A)$ and misses the block with probability $\mathbb P_1(\mathbf Y\notin A) = 1- \mathbb P_1 (\mathbf Y\in A)$.

Under $H_1$, the coordinating group $\Sstar$ is drawn uniformly from $\mathcal S_k$, so $\mathbb P_1$ is the average, over all $S\in\mathcal S_k$, of the distribution of $\mathbf Y$ when $\Sstar=S$. The likelihood ratio $L=d\mathbb P_1/d\mathbb P_0$ of $H_1$ against $H_0$ satisfies $\E_0[L]=1$. For every test, that is, for every set $A$,
\begin{align*}
\mathbb P_0(\mathbf Y\in A)+\mathbb P_1(\mathbf Y\notin A)
&=\\
&\mathbb P_0(\mathbf Y\in A)+1-\mathbb P_1(\mathbf Y\in A)\\
&\ge 1-\|\mathbb P_1-\mathbb P_0\|_{\mathrm{TV}}\\
&\ge 1-\tfrac12\sqrt{\E_0[L^2]-1}.
\end{align*}
The first inequality holds because $\mathbb P_1(\mathbf Y\in A)-\mathbb P_0(\mathbf Y\in A)\le\|\mathbb P_1-\mathbb P_0\|_{\mathrm{TV}}$ for every set $A$. The second holds because $\|\mathbb P_1-\mathbb P_0\|_{\mathrm{TV}}=\tfrac12\E_0|L-1|\le\tfrac12\sqrt{\E_0[(L-1)^2]}$ by the Cauchy--Schwarz inequality, and $\E_0[(L-1)^2]=\E_0[L^2]-1$ \citep{tsybakov2009introduction}. Since $\mathbb P_1$ averages over $\Sstar$, the bound holds for the error averaged over the possible groups, and therefore also for the largest error over all groups, which is at least the average.

Let $\ell=dP/dQ$ be the likelihood ratio of a single score of the block against a null score; its logarithm is Equation~\eqref{eq:llr} (Appendix~\ref{app:score}). If the group $\Sstar=S$ were known, the likelihood ratio of the whole matrix would be $\prod_{i<j\in S}\ell(Y_{ij})$, since the entries outside $S$ have the same distribution $Q$ under both hypotheses. Under $H_1$ the group is uniform on $\mathcal S_k$, so $L$ is the average of these products:
\[
L=\binom nk^{-1}\sum_{S\in\mathcal S_k}\ \prod_{i<j\in S}\ell(Y_{ij}).
\]

\emph{Second moment.} Squaring $L$ gives a double sum over two groups $S$ and $S'$, which we can read as two independent groups drawn uniformly from $\mathcal S_k$:

\[
\E_0[L^2]=\binom nk^{-2}\sum_{S,S'\in\mathcal S_k}\E_0\Big[\prod_{i<j\in S}\ell(Y_{ij})\prod_{i<j\in S'}\ell(Y_{ij})\Big].
\]

Under $H_0$ the scores are independent, so the expectation is a product over pairs. A pair in neither group contributes $1$, and a pair in exactly one group contributes $\E_0[\ell]=1$. A pair in both groups has both accounts in $S\cap S'$ and contributes

\begin{equation}
\begin{split}
\E_0[\ell^2]=\int_0^1\big(\lambda x^{\lambda-1}\big)^2dx=\\
\frac{\lambda^2}{2\lambda-1}=\frac1{1-\mu^2}=1+\chi^2(P\Vert Q),
\label{eq:rdef}
\end{split}
\end{equation}
with $\lambda=1/(1+\mu)$. The integral is over the $p$-value $x$, which is uniform on $(0,1)$ under $H_0$. This gives $\E_0[\ell^2]$ because the score is a one-to-one function of the $p$-value, so $\ell$ has the same value whether it is computed from $Y_{ij}$ or from $p_{ij}$ (Appendix~\ref{app:score}). The integral is finite only for $\lambda>1/2$, that is, for $\mu<1$. If the two groups share $K=|S\cap S'|$ accounts, exactly $\binom K2$ pairs are in both groups. For two independent uniform groups, $K$ has the hypergeometric distribution $\Pr(K=j)=\binom kj\binom{n-k}{k-j}\big/\binom nk$. Hence

\begin{equation}
\E_0[L^2]=\E\Big[(1-\mu^2)^{-\binom K2}\Big]=\sum_{j=0}^{k}\frac{\binom kj\binom{n-k}{k-j}}{\binom nk}\,(1-\mu^2)^{-\binom j2}.
\label{eq:secmom-exact}
\end{equation}
Every term increases with $\mu$. At $\mu=0$ the sum equals $1$, and as $\mu\to1$ it tends to infinity, because of the terms with $j\ge2$. So, for every $\gamma>1$, there is exactly one $\mu$ at which $\E_0[L^2]=\gamma$. We define $\mu_{\mathrm{IT}}$ by $\E_0[L^2]=e$. For $\mu<\mu_{\mathrm{IT}}$, the bound above gives a total error probability of at least $1-\tfrac12\sqrt{e-1}=0.34$ for every test; this holds at finite $(n,k)$ when $\mu_{\mathrm{IT}}$ is computed from Equation~\eqref{eq:secmom-exact}.

\emph{Closed forms.} For small groups ($k\ll\sqrt n$) the overlap $K$ has mean $\E[K]=k^2/n\to0$. The terms $j=0$ and $j=1$ of Equation~\eqref{eq:secmom-exact} have the factor $(1-\mu^2)^0=1$ and sum to $1+o(1)$; the terms with small $j\ge2$ vanish, and the term that remains is $j=k$. This term equals $(1-\mu^2)^{-m}/\binom nk$, with $m=\binom k2$, because the two groups coincide ($K=k$) with probability $\binom nk^{-1}$ and then share all $m$ pairs. It is of order $1$ when $m\big(-\log(1-\mu^2)\big)=\log\binom nk+O(1)$. As $\mu\to0$, $-\log(1-\mu^2)=\mu^2(1+o(1))$, and $\log\binom nk=k\log(n/k)\,(1+o(1))$ because $k\log(n/k)\le\log\binom nk\le k\log(n/k)+k$. The condition becomes $m\mu^2=k\log(n/k)\,(1+o(1))$, and with $m=k(k-1)/2$,

$$\mu=\sqrt{\frac{2\log(n/k)}{k-1}}\,\big(1+o(1)\big).$$

The value of $\gamma$ changes $\log\binom nk$ only by $O(1)$, so it does not affect the leading term of $\mu$

For large groups ($k\gg\sqrt n$) the overlap $K$ has mean $\kappa=k^2/n\to\infty$ and standard deviation at most $\sqrt\kappa$, so $K/\kappa\to1$ and we can replace $K$ by $\kappa$. The expectation in Equation~\eqref{eq:secmom-exact} is $\E\big[\exp\big(g(K)\big)\big]$, with $g(K)=-\binom K2\log(1-\mu^2)$. As $\mu\to0$, $-\log(1-\mu^2)/\mu^2\to1$, and as $K\to\infty$, $\binom K2\big/(K^2/2)\to1$; together with $K/\kappa\to1$, this gives $g(K)\big/(\mu^2\kappa^2/2)\to1$. So $\E_0[L^2]$ is approximately $\exp(\mu^2\kappa^2/2)$, which equals $\gamma$ at $\mu=\sqrt{2\log\gamma}/\kappa=\sqrt{2\log\gamma}\,n/k^2$. The approximation holds at this $\mu$: since $K$ differs from $\kappa$ by about $\sqrt\kappa$, replacing $K$ by $\kappa$ changes $g$ by about $\mu^2\kappa\sqrt\kappa\approx1/\sqrt\kappa$, which tends to $0$. For $\gamma=e$ the value is $\sqrt2\,n/k^2$, whose ratio to $\mu_{\mathrm{sum}}=\sqrt{2n(n-1)}/(k(k-1))$ tends to $1$.

$\E_0[L^2]$ exceeds $e$ as soon as either the term $j=k$ or the bulk of the overlap distribution does, so to leading order
\[
\mu_{\mathrm{IT}}=\min\Big\{\sqrt{2\log(n/k)/(k-1)},\ \sqrt2\,n/k^2\Big\}.
\]

This minimum is an approximation of the exact value of $\mu_{\mathrm{IT}}$, the numerical solution of $\E_0[L^2]=e$ with Equation~\eqref{eq:secmom-exact}, and the ratio of the two tends to $1$ as $n\to\infty$. On the $199$ layers of Section~\ref{sec:results}, the exact value is between $0.54$ and $1.10$ times the closed form (median $0.99$); the largest differences occur for groups with fewer than $10$ accounts.

\emph{Gap to the scan.} Where $\mu_{\mathrm{IT}}$ equals its first term, that is, for $k$ below about $n^{2/3}$, $\mu_{\mathrm{IT}}$ is a factor $\sqrt2$ below $\mu_{\mathrm{scan}}$ (Appendix~\ref{app:scanthresh}). The smallest signal at which some test detects the block may lie above $\mu_{\mathrm{IT}}$, because rare large overlaps $K$ can inflate $\E_0[L^2]$. For Gaussian entries, a truncated second moment that removes these overlaps closes the gap to the scan \citep{butucea2013detection}; this has not been done for exponential entries. For $\mu_{\mathrm{IT}}\le\mu<\mu_{\mathrm{scan}}$ it is therefore not proved that any test detects the block, nor that none does.

\subsection{Sum boundary}
\label{app:sum}

Our sum test $T^{(l)}_{\mathrm{sum}}$ is the linear test of \citep{butucea2013detection}, applied to the $N=\binom n2$ scores above the diagonal. As in \citep{butucea2013detection}, we divide $T^{(l)}_{\mathrm{sum}}$ by its standard deviation under $H_0$. The scores are independent by assumption, so with $m=\binom k2$, Equation~\eqref{eq:oracle-h1-exact} applied to the block and to the remaining $N-m$ pairs gives
\begin{align*}
&\E_0[T^{(l)}_{\mathrm{sum}}]=0,\quad \Var_0(T^{(l)}_{\mathrm{sum}})=N,\\
&\E_1[T^{(l)}_{\mathrm{sum}}]=m\mu,\quad \Var_1(T^{(l)}_{\mathrm{sum}})=N+m\mu(2+\mu).
\end{align*}
The normalised sum $T^{(l)}_{\mathrm{sum}}/\sqrt N$ has mean $0$ and variance $1$ under $H_0$, and mean $\delta=m\mu/\sqrt N$ and variance $1+2\delta/\sqrt N+\delta^2/m$ under $H_1$. 

By Chebyshev's inequality, the test $T^{(l)}_{\mathrm{sum}}>z\sqrt N$ has type~I error at most $1/z^2$ and, for $z<\delta$, type~II error at most $\big(1+2\delta/\sqrt N+\delta^2/m\big)/(\delta-z)^2$. For $z=\sqrt\delta$, both bounds tend to $0$ when $\delta\to\infty$, as for Gaussian entries \citep{butucea2013detection}.

In \citep{butucea2013detection}, the boundary of the linear test is where the normalised mean shift is approximately one. Following this, $\mu_{\mathrm{sum}}$ is the signal strength at which $\delta=1$:
\begin{align*}
\mu_{\mathrm{sum}}=\frac{\sqrt N}{m}=\frac{\sqrt{2n(n-1)}}{k(k-1)}.
\end{align*}
This value also equals the large-group term of $\mu_{\mathrm{IT}}$ with $\gamma=e$ (Appendix~\ref{app:it-bound}).

For $k\ge\sqrt n$, the order of $\mu_{\mathrm{sum}}$ is also the computational boundary of \citep{brennan2019universality}, which is $\dskl\asymp n^2/k^4$, up to factors that grow slower than any power of $n$.
With $\dskl=\mu^2/(1+\mu)$ (Equation~\eqref{eq:dskl-chi}), this is $\mu\asymp n/k^2$ as $\mu\to0$. Below this order no polynomial-time test detects the block under the planted clique hypothesis, and above it a polynomial-time test does (Appendix~\ref{app:threeregimes}).

\subsection{Spectral boundary}
\label{app:spectral}

Following the spectral-edge and matrix-deformation results
of \cite{bai1988necessary,knowles2013isotropic}, we derive the
spectral detection threshold stated in Section~\ref{sec:threeregimes} and examine
the effect of the larger variance of entries within the
planted block.

The spectral boundary follows from the largest eigenvalue of a random matrix with a low-rank perturbation \citep{bai1988necessary, knowles2013isotropic}. Write $\mathbf Y=\E[\mathbf Y]+(\mathbf Y-\E[\mathbf Y])$. Under $H_1$, $\E[\mathbf Y]$ has the entry $\mu$ on the pairs of $\Sstar$ and $0$ elsewhere, so its largest eigenvalue is $\mu(k-1)$. Let $\theta=\mu(k-1)/\sqrt n$. Outside the block, the entries of $\mathbf Y-\E[\mathbf Y]$ are independent, with distribution $\mathrm{Exp}(1)-1$, mean $0$ and variance $1$ (Appendix~\ref{app:score}). Under $H_0$, the largest eigenvalue of $\mathbf Y/\sqrt n$ tends to $2$ \citep{bai1988necessary}. Under $H_1$, it tends to $\theta+1/\theta$ if $\theta>1$ and to $2$ if $\theta\le1$ \citep{knowles2013isotropic}. The test that rejects $H_0$ when the largest eigenvalue of $\mathbf Y/\sqrt n$ exceeds $2+\varepsilon$ therefore detects the block if $\theta+1/\theta>2+\varepsilon$, and for no $\varepsilon>0$ if $\theta\le1$, which gives $\mu_{\mathrm{spec}}=\sqrt n/(k-1)$.

The block entries of $\mathbf Y-\E[\mathbf Y]$ have variance $(1+\mu)^2$ instead of $1$, because a centred block score has the distribution of $1+\mu$ times a centred score under $H_0$ (Appendix~\ref{app:score}). So $\mathbf Y-\E[\mathbf Y]$ has the distribution of a matrix of centred scores under $H_0$ plus $\mu$ times the same matrix with all entries outside the pairs of $\Sstar$ set to $0$. The nonzero part of this second matrix is a $k\times k$ matrix of centred scores under $H_0$, so its eigenvalues lie between about $-2\sqrt k$ and $2\sqrt k$ \citep{bai1988necessary}. By Weyl's inequality, adding a matrix changes each eigenvalue by at most the largest absolute eigenvalue of the added matrix \citep{fulton2000eigenvalues}. The largest eigenvalue of $\mathbf Y/\sqrt n$ therefore changes by at most about $2\mu\sqrt k/\sqrt n=2\theta/\sqrt k$, which tends to $0$ as $k\to\infty$, and the limits above also hold with the larger variance of the block entries.

\subsection{Computational boundary}
\label{app:threeregimes}

We verify the conditions under which the reduction from the planted clique problem of \citet{brennan2019universality} applies to our scores. Under $P$, $\log\ell=\mu E-\log(1+\mu)$ with $E\sim\mathrm{Exp}(1)$ (Appendix~\ref{app:score}), so
\begin{equation}
\begin{split}
\mathrm{KL}(P\Vert Q)=\mu-\log(1+\mu)=\tfrac{\mu^2}{2}\big(1+O(\mu)\big),\\
\dskl=\frac{\mu^2}{1+\mu},
\label{eq:dskl-chi}
\end{split}
\end{equation}
and $\Pr_P(\log\ell\ge\tau)$ decays with the exponent $I_P(\tau)=u-1-\log u$, where $u=(\tau+\log(1+\mu))/\mu$.

Besides the condition on $\mathrm{KL}(P\Vert Q)$, the lower bound of \citep{brennan2019universality} needs two more. First, $(P,Q)$ must be a computable pair: $Q$ can be sampled and $\ell$ evaluated in polynomial time, $\E_Q[\ell]=\E_P[\ell^{-1}]=1$, and both KL divergences are finite. This holds because $P$ and $Q$ are exponential with the same support. Second, large values of $\log\ell$ must be rare under $P$, because the reduction from the planted clique problem in \citep{brennan2019universality} is accurate only on entries whose $\log\ell$ stays below a threshold $\tau$. This is measured  with the Chernoff exponent $I_P(\tau)=\sup_t\{t\tau-\log\E_P[\ell^t]\}$, which gives $\Pr_P(\log\ell\ge\tau)\le e^{-I_P(\tau)}$ for $\tau\ge\mathrm{KL}(P\Vert Q)$ \citep{brennan2019universality}. For our scores, $I_P(\tau)=u-1-\log u$ with $u=(\tau+\log(1+\mu))/\mu$.

For $k\ge\sqrt n$, the second condition is that some threshold $\tau$ satisfies $\tau\ge\mathrm{KL}(P\Vert Q)$, $\tau k^4/n^2\to0$, and $I_P(\tau)/(\tau\log n)\to\infty$. If $\mu=o(\mu_{\mathrm{sum}})$, that is, $\alpha>4\beta-2$, the threshold $\tau=\mu^a$ with $1<a<2$ and $a\alpha/2>4\beta-2$ satisfies all three: $\mathrm{KL}(P\Vert Q)\approx\mu^2/2\le\mu^a$; $\tau k^4/n^2=n^{4\beta-2-a\alpha/2}\to0$; and $I_P(\tau)\approx\mu^{2a-2}/2$, so $I_P(\tau)/(\tau\log n)\approx\mu^{a-2}/(2\log n)\to\infty$. For $k<\sqrt n$, the second condition is $I_P(\log\rho^{-1})\ge2\log n$, with $\rho$ the edge density of the planted clique hypothesis. This needs $u \gtrsim 2\log n$, that is, $\mu=O(1/\log n)$.

\subsection{Coordinates of the phase diagram}
\label{app:coords}

The phase diagram for submatrix detection in \citet{brennan2019universality} has two coordinates: the size of the block, $k=n^{\beta}$, and the signal, $\dskl=n^{-\alpha}$, both up to factors that grow slower than any power of $n$. In these coordinates, detection is impossible for $\alpha>\max(\beta,4\beta-2)$, possible in polynomial time for $\alpha<\max(0,4\beta-2)$, and, under the planted clique hypothesis, hard in between.

We use the same coordinates, with the signal $\mu^2$ in place of $\dskl$, since $\dskl=\mu^2/(1+\mu)$ (Equation~\eqref{eq:dskl-chi}). The exponent of $\dskl$ differs from $\alpha$ by $\log(1+\mu)/\log n\le\log2/\log n$ for $\mu\le1$, which tends to $0$ for $n \to \infty$. Each boundary has the form $\mu^2=c\,L\,n^{a}k^{-b}$, with $k-1\approx k$, a constant $c$, a factor $L$ equal to $1$ or $\log(n/k)$, and exponents $a$ and $b$. With $k=n^\beta$ and $\mu^2=n^{-\alpha}$, the boundary is $\alpha=b\beta-a-\log(cL)/\log n$. The last term is $O(\log\log n/\log n)$ and tends to $0$, so as $n\to\infty$ the boundary becomes the line $\alpha=b\beta-a$. The scan boundary and the small-group term of $\mu_{\mathrm{IT}}$ have $\mu^2\propto\log(n/k)/k$, which gives $\alpha=\beta$; the sum boundary and the large-group term of $\mu_{\mathrm{IT}}$ have $\mu^2\approx2n^2/k^4$, which gives $\alpha=4\beta-2$, as in \citep{brennan2019universality}. The spectral boundary has $\mu^2\approx n/k^2$, which gives $\alpha=2\beta-1$. This line is not in \citep{brennan2019universality}, where, for $k<\sqrt n$, detection is possible in polynomial time already for $\alpha<0$. Layers with $k<\sqrt n$ and $2\beta-1<\alpha<0$ are therefore hard for the sum and spectral tests.

\end{document}